\documentclass[]{aa}
\usepackage{graphicx}
\begin{document}
\title{The star formation history of the LSB Galaxy UGC 5889
\thanks{Based on observations collected with Hubble Space Telescope}}
%\titlerunning{ HST Photometry of UGC 5889}
\titlerunning{ Star formation history of UGC 5889}

\author{A. Vallenari$^1$, L. Schmidtobreick$^2$, D.J. Bomans$^3$ }

\institute{
$^1$Astronomical Observatory of Padova, Vicolo dell' 
              Osservatorio 5, 35122 Padua, Italy \\
$^2$European Southern Observatory, Casilla 19001, Santiago 19, Chile\\ 
$^3$Astronomisches Institut der Ruhruniversit\"at Bochum, 
44780 Bochum, Germany \\
email vallenari@astrpd.pd.astro.it}

\offprints {A. Vallenari}

\date{Received April 2004, Accepted January 2005}

\abstract{
We present HST photometry of the LSB galaxy UGC 5889 and 
derive its recent star formation history. In the last 200 Myr 
the star formation  proceeded
in modest bursts at a rate of the order of 10$^{-2}$-10$^{-3}$
M$_{\sun}$/yr,  
with periods of extremely low SFR or even quiescence. 
The rate derived from the present study
for the last 20 Myr  is in agreement with the
H$_\alpha$ emission from the galaxy.
The presence of a consistent population older than 200 Myr is suggested by the data.
However,
observational errors and completeness correction prevent any firm conclusion
on the  oldest age. The total mass of stars is of the order of 5.5 $\times 10^7$ M$_{\sun}$. 
Even if the recent episodes of star formation have heated the gas and carved
a hole in the disk,  blow-away of the gas is unlikely to occur.
%UGC~5889 is then probably not  
%a transition object between an LSB and a dwarf elliptical.
%The present result points in favor of  the hypothesis that the blue colour
%of the LSBs is not due to a pure metallicity effect, 
%but is depending on the fact that star formation process is going on in
%sporadic episodes.

    }

\maketitle

\section{Introduction }

%To derive the stellar content and
% evolutionary path of LSB galaxies is of fundamental importance towards
%understanding the formation and evolution history of all galaxies, since
%dwarf galaxies are considered as building blocks in the hierarchical scenario
It is well known that low surface brightness  (LSB) galaxies
 contribute significantly to the
total galaxy counts (see e.g. Marzke \& da Costa, \cite{marz+97}) and might 
even constitute the majority of galaxies (O'Neil \& Bothun \cite{onei+00}). 
LSB galaxies show a large variety of
properties, ranging from the blue to the red end of galaxy colours (O'Neil
et al. 1997, 2000) and 
consisting of dwarfs as well as of massive systems (Sprayberry et al. 1995).
Concerning their metal content, the vast majority of LSB galaxies show low metal content 
(de Block \&van der Hulst 1998), although some red LSB galaxies having solar 
metallicity have been recently detected (Bergmann et al. 2003).
Among these diverse systems,
we focus on one of the most common types, the blue LSB (BLSB) galaxies
(see the review by Impey \& Bothun \cite{impe+97} and references therein), late-type disk galaxies having B surface brightness $> 23$ mag arcsec$^{-2}$ where colours of 
  (B-V)$<0.6$ are found
(O'Neil et al. \cite{onei+97}).
These blue colours are  puzzling primarily because of the restrictions they put
on the star formation history of LSB galaxies. 
%Late type high surface brightness
%spirals normally have
%similar colours (Huchra \cite{huch77}). 
%However, the star formation rate 
%of LSB is on average a factor of ten lower that late spirals (McGaugh \& Blok \cite{mcga+97}).
Even if they present low surface gas density,
BLSB are among the most gas-rich galaxies at a given total luminosity.
The high gas fraction of  BLSB galaxies, together with their low metal
content, suggest that they are relatively unevolved objects.
This can be interpreted in various ways. For instance 
BLSB galaxies can have very low
present and past star formation rates (SFR): 
in these systems the main phase of star
formation is still to occur.
Alternatively, these galaxies can have a stellar population similar to those of
high surface brightness spirals, but their luminosity can be dominated
by a young population.
%Their blue colours are  puzzling primarily because of the restrictions they put
%on the star formation history of LSB. Late type spirals normally have
%similar colours (Huchra \cite{huch77}). However, the star formation rate 
%of LSB is on average a factor of ten lower that late spirals (McGaugh \& Blok 
%\cite{mcga+97}).
%One explanation can be found in a  
Until now,
the investigation of the stellar content of LSB galaxies has mainly
 been based on the
discussion of the integrated colours. Many studies
 are consistent with a  scenario where BLSB galaxies have on average a 
constant or increasing SFR with a ratio of young to old stars
larger than found in high brightness spirals
%has been proceeding in frequent small amplitude busts 
(van den Hoek et al. 2000, Knezek 1993, Bell \& de Jong 1999). 
%A combination of
% low metal content
%although
%  a comparison with the colours of the metal poor Galactic globular
%cluster suggests that 
%very blue LSBs present bluer colours than
% metal poor Galactic globular clusters
%(Knezek et al. \cite{knez+99}).
%Alternatively,
% the simplest idea is that blue colours are
%together  recent SF episodes can be responsible of the very blue colours
%(Knezek et al. \cite{knez+99}, van den Hoek et al 2000).
%Theoretical models predict that  100 Myr after the
%SF episode the color of the population has faded of $\Delta$(U-B)$\sim -0.2$
%(O'Neil et al. \cite{onei+98}). 
%Because of the low
%surface brightness of the underlying population, only a small
%fraction of young stars is needed to make the colour significantly blue.
%Integrated colours are consistent with a  scenario where the SFR
%has been proceeding in frequent small amplitude busts (van den Hoek et al 2000). 
However, different opinions can be found in the literature,
either suggesting that the  SFR
can be  proceeding in frequent, small amplitude bursts 
(van den Hoek et al. 2000)
or that the bulk of the population is older than 5-7 Gyr
(Padoan et al. 1997).
Finally, it is far from being understood how stars form in a low density, low metal content environment and what is the relation between SFR and other
properties of LSB galaxies (i.e. the angular momentum) (see among  others O'Neil et al. 2000, Boissier et al. 2003, Dalcanton, Spergel \& Summers 1997).

% This  to the 
%well known age-metallicity degeneracy affecting the interpretation
%of  broad-band colours.   

%However theoretical models predict that  100 Myr after the
%Sf episode the color of the population has faded to (U-B)$\sim 0$
%(O'Neil et al. \cite{onei+98}).
%The low  present day SFR requires that the star 
%formation occurs in sporadic bursts (O'Neil et al. \cite{onei+98}) . 
%Up to now
%the investigation of the stellar content of LSBs has mainly
% been based on the
%discussion of the integrated colours.
In this paper we focus on the study
of a resolved LSB galaxy: UGC 5889 (NGC 3377A).
This dwarf galaxy is a member of the Leo I group.
Due to its similar redshift
and resulting proximity -- only 20\,Kpc in projected distance 
(Sandage \& Hoffman, \cite{sand+91}),
it is thought to be a
companion of the large elliptical NGC 3377.
With M$_{H_I}/$L$_B \sim 0.3$ M$_\odot/$ L$_\odot$, it is gas rich
(Knezek et al. \cite{knez+99}).
It is classified as a low surface brightness  dwarf
having an average surface brightness inside the Holmberg radius of B=23.97 mag arcsec $^{-2}$.
Moderately blue colours (U-B)$= 0.06 \pm 0.05$, and (B-V)$= 0.64 \pm 0.06$ 
are given by Prugniel \& Heraudeau (\cite{prug+98}).

Base on its optical morphology, Knezek et al. (\cite{knez+99}) suggested the 
presence of a spiral structure classifying the galaxy as SABm. This has recently been supported by
high resolution spectroscopic H\,I--measurements of 
Huchtmeier et al. (\cite{huch+03}), 
who find a double peaked profile for this galaxy which is typical for spirals. 
%However, the young star groupings (YSGs) that have been discovered by 
%Vicari et al. (\cite{vica+02}) inside UGC\,5889 do not follow any 
%obvious spiral pattern, but homogeneously fill the whole optical area 
%of the galaxy. 
The H\,I intensity maps of
Simpson \& Gottesman (\cite{simp+00})
show a ring--shaped distribution of the gas which is surrounding the bulk of the optical body and  is correlated with the young star
groups discovered by
Vicari et al.(\cite{vica+02}).
% which could also be interpreted as rudimentary
%spiral arms. 

Knezek et al. (\cite{knez+99}) also measured the H$_\alpha$ luminosity of
UGC\,5889 and found a relatively low value of  
L$_{H_\alpha} \sim 2.6 \times 10^{38}$ ergs s$^{-1}$, which is
about ten times lower than the average H$_\alpha$ luminosity in the sample
of quiescent dwarf spiral galaxies studied by van~Zee et al. (\cite{vanz+97}). 
Thus, despite the dense environment where it is located,
 UGC 5889 appears to be a very quiet dwarf galaxy.  
Knezek et al. (\cite{knez+99}) discuss the possibility that it
might represent a true transition object between  
late type dwarf galaxies and low SFR dwarf 
ellipticals  as suggested by 
Sandage \& Hoffman (\cite{sand+91}). However
they reject the hypothesis, since 
%However, 
with its large gas reservoir
%with M$_{H_I}/$L$_B \sim 0.3$ 
UGC\,5889 
 can continue to form stars at the present  SFR for much longer than
the Hubble time. 

%Knezek et al. therefore argue that it is most unlikely 
%for UGC\,5889 to evolve into a
%dwarf elliptical.
%In this paper we verify this  interpretation on the basis
%of the SFR derived from the CMD.
% This galaxy  does not represent a "missing link".
 
% This latter being the case, future star formation bursts might 
%allow to get rid of a large amount of gas. 
%UGC\,5889 evolves indeed into a dwarf elliptical.  
%To actually judge the evolutionary progression of any dwarf galaxy
%the knowledge of its star formation history is mandatory, neither
%the optical morphology nor the gas content is sufficient to determine
%its evolutionary phase.

%In this paper we focus on the determination of the star formation history
%of UGC 5889 on the basis of HST archive data. 

The star formation history of UGC 5889 is unknown. It is still not clear
whether the current SFR represents the mean
level over the galaxy lifetime or, whether  the
star formation of this galaxy has occurred
 at a different rate in the past. This paper deals with the
determination of the SFR and the SFR history from the analysis
of colour--magnitude diagrams (CMDs).
The plan of this paper is as follows:
 in Section \ref{obs}, we discuss the observations, the data
reduction and we present the observed CMD.
In Section \ref{distance}, we derive the distance modulus, in section
\ref{cmd}, the basic CMD features. In 
Section \ref {sfh} we determine the star formation history of the galaxy
and we discuss the feedback from star formation.
In Section \ref {comparison}, we compare the SFR of UGC~5889 with other
LSB galaxies.
Conclusions are drawn in Section \ref{conclusions}.

\section{Observation and reduction}
\label{obs}
\begin{figure*}[t]
%%\centerline{\resizebox{18.0cm}{!}{\includegraphics{ugc5889.eps}}}
\centerline{\resizebox{18.0cm}{!}{\includegraphics{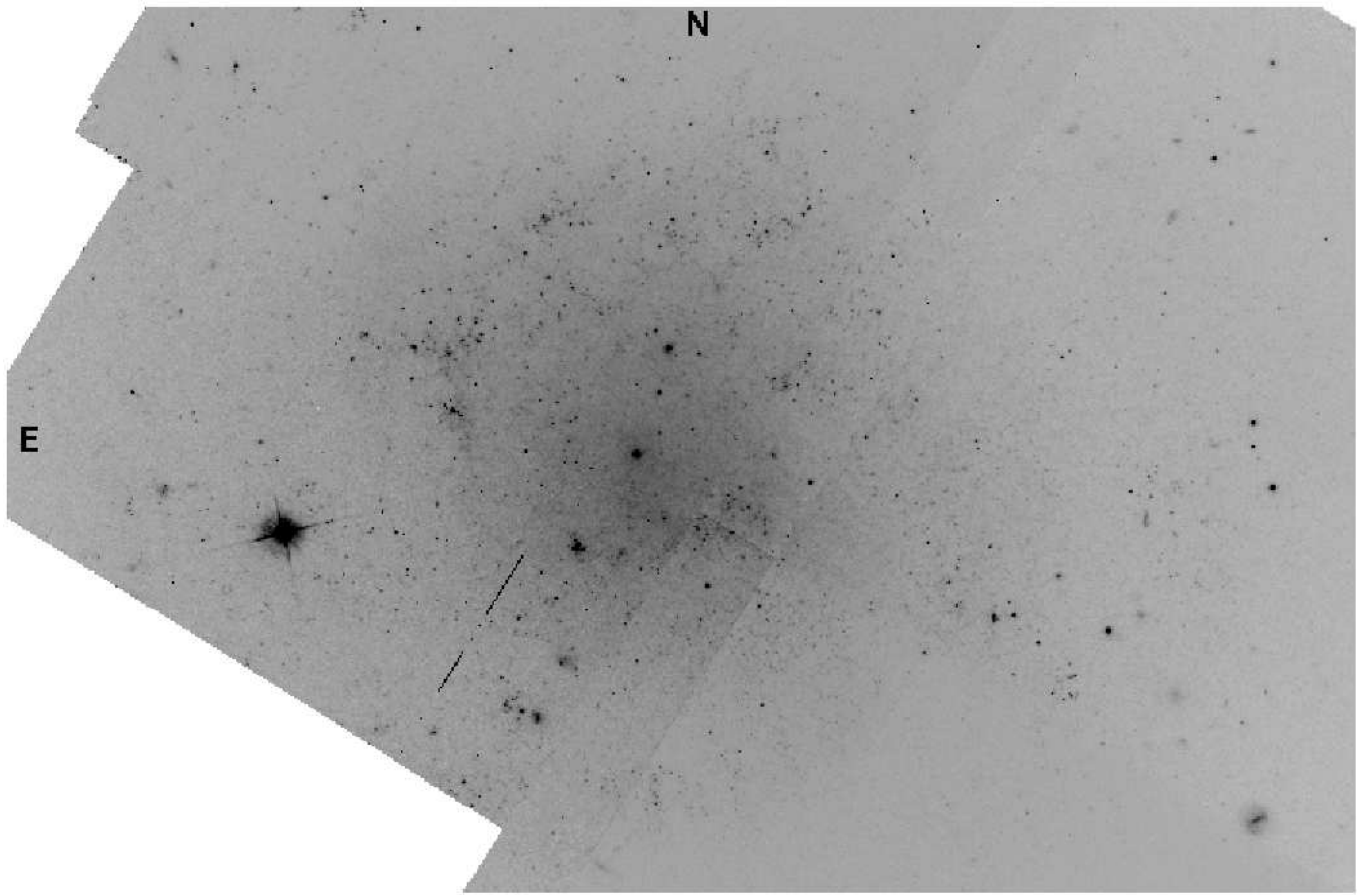}}}
\caption{\label{image} Composite  image of the central area of the 
galaxy UGC 5889.}
\end{figure*}

We use HST archive data of
UGC 5889 taken in 1995 with WFPC2 at two different 
position angles and locations. 
In total, 21 partially overlapping F555W and four F814W frames
are used including long and short exposed frames and yielding a 
total exposure time of 6000\,s in F814W and 25300\,s in F555W.

A composite image of the galaxy in the F555W filter is presented in
Fig.\ref{image}. It shows the high spatial resolution, which allowed the
photometric measurement of individual stars.

The data have been reduced using the stand--alone version of
DAOPHOT including ALLFRAME and following the ALLFRAME cookbook
(Turner \cite{turn96}). 
To de-select suspected clusters and foreground galaxies we excluded the
objects with FWHM larger than 2 pixels using the ALLFRAME parameter
'sharp'. In total, we derive photometry for 4843 stars in the galaxy
spanning the F814W magnitude range from  20$^{\rm m}$ to fainter then
27$^{\rm m}$.

\begin{figure}[bt]
%\vspace{-8mm}
%\special{psfile=figcomp2b.ps
%         hoffset=-30 voffset=00 hscale=37 vscale=37 angle=-90}
%\vspace{53mm}
%%{\resizebox{!}{8.8cm}{\includegraphics{comple3.ps}}}
{\resizebox{!}{8.8cm}{\includegraphics{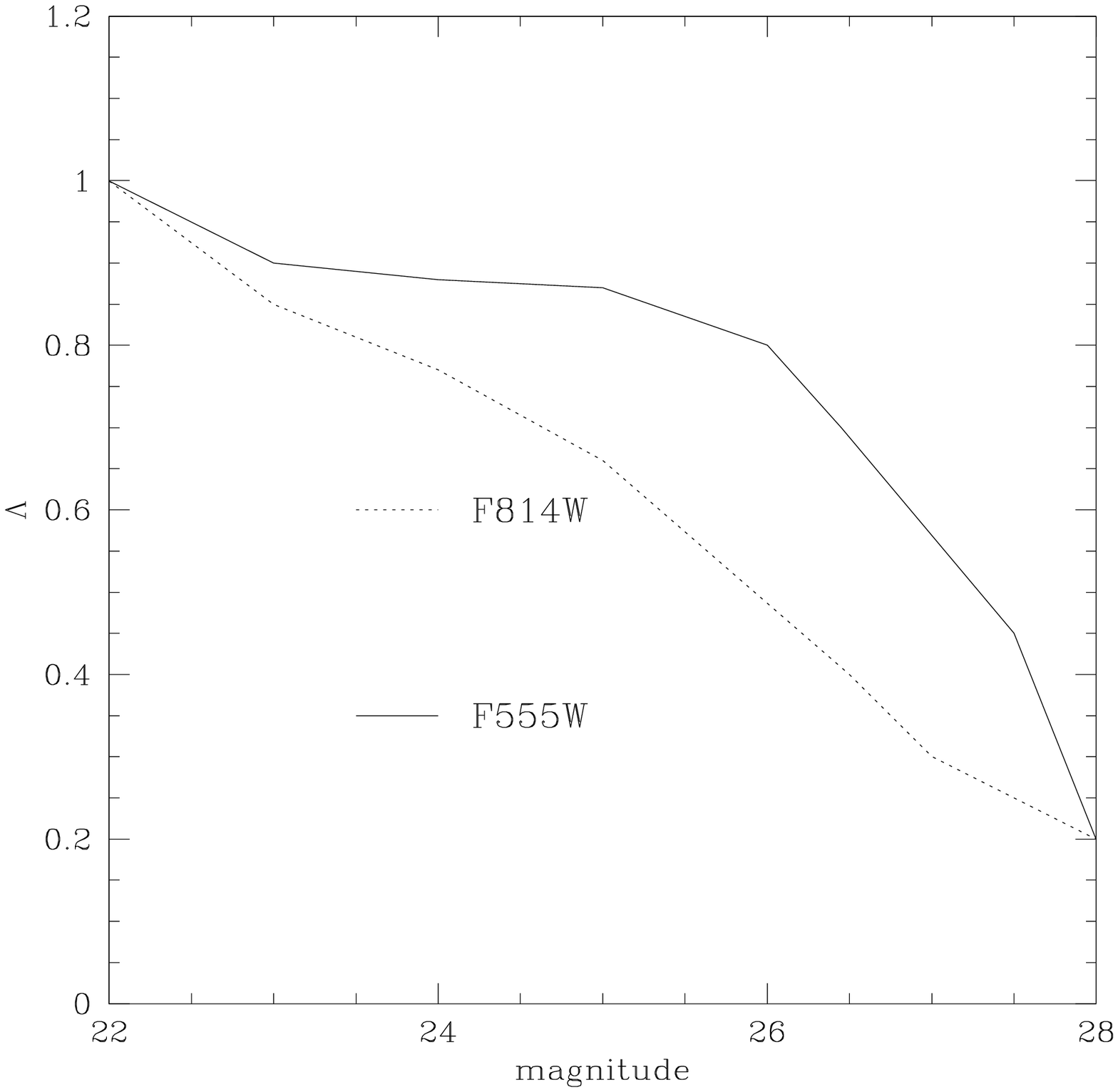}}}
\caption{\label{comple} Completeness factors $\Lambda$ 
for F555W and F814W magnitudes}
\end{figure}

%i%The resulting photometric errors are listed in Table \ref{errors}.
%The  photometric errors are $<$0.1, 0.2, 0.30, 0.40, 0.5 for F555W magnitudes
%$<$24,25,26,27,28 respectively.
The photometric calibration has been done following the prescriptions by
Dolphin (\cite{dolph}).
We estimate the  error on the zero point as 0.1 mag.
The completeness factors $\Lambda$ are calculated as usual by means 
of artificial star experiments and   are plotted
in Fig. \ref{comple}. The data are complete at 40\% level for magnitudes
brighter than  F555W $<$27.5 or F814W $<$26.5.
 Artificial star experiments are used as well to derive
the errors on the magnitudes due both to the Poisson noise and crowding.
These errors are listed in Table \ref{errors}.  Photometric errors due only
to the Poisson statistics on the noise are estimated by ALLFRAME. As expected, they
are smaller, being $<$0.1, 0.14, 0.22, 0.30 for F555W magnitudes $<$25,26,27,27.5, respectively. The analogous photometric errors
on F814W magnitude are  $<$0.1, 0.12, 0.24 for F814W $=$ 24,25,26 respectively.

\begin{table}
\caption{\label{errors} The photometric errors as a function of the
  magnitude.}
\begin{tabular}{l c c c c c }
\hline
 \noalign{\smallskip}
F555W [mag] & $<24$ & $25$ & $26$ & $27$ & $27.5$ \\
 \noalign{\smallskip}
\hline
 \noalign{\smallskip}
$\sigma _{\rm phot}$ [mag] & $<0.1$ & 0.1 & 0.2 & 0.3 & 0.45 \\
 \noalign{\smallskip}
\hline
\hline
 \noalign{\smallskip}
F814W [mag] & $<24$ & $25$ & $26$ &  &  \\
 \noalign{\smallskip}
\hline
 \noalign{\smallskip}
$\sigma _{\rm phot}$ [mag] & $<0.1$ & 0.2 & 0.3 & &  \\
 \noalign{\smallskip}
\hline
\end{tabular}
\end{table}

%In the following we discuss the 
%6 F555W frames and 2 F814W frames taken between ***.
%The total exposure time in F555W is 7200s and in F814W it is 3000s.
%We used DAOPHOT/ALLSTAR inside the IRAF enviroment.

%\section{The data}
%\label{data}

\begin{figure}[bt]
%\vspace{-8mm}
%\special{psfile=figcomp2b.ps
%         hoffset=-30 voffset=00 hscale=37 vscale=37 angle=-90}
%\vspace{53mm}
%{\resizebox{9.0cm}{!}{\includegraphics{u5889cmd0.ps}}}
%%{\resizebox{9.0cm}{!}{\includegraphics{fig3_i.ps}}}
{\resizebox{9.0cm}{!}{\includegraphics{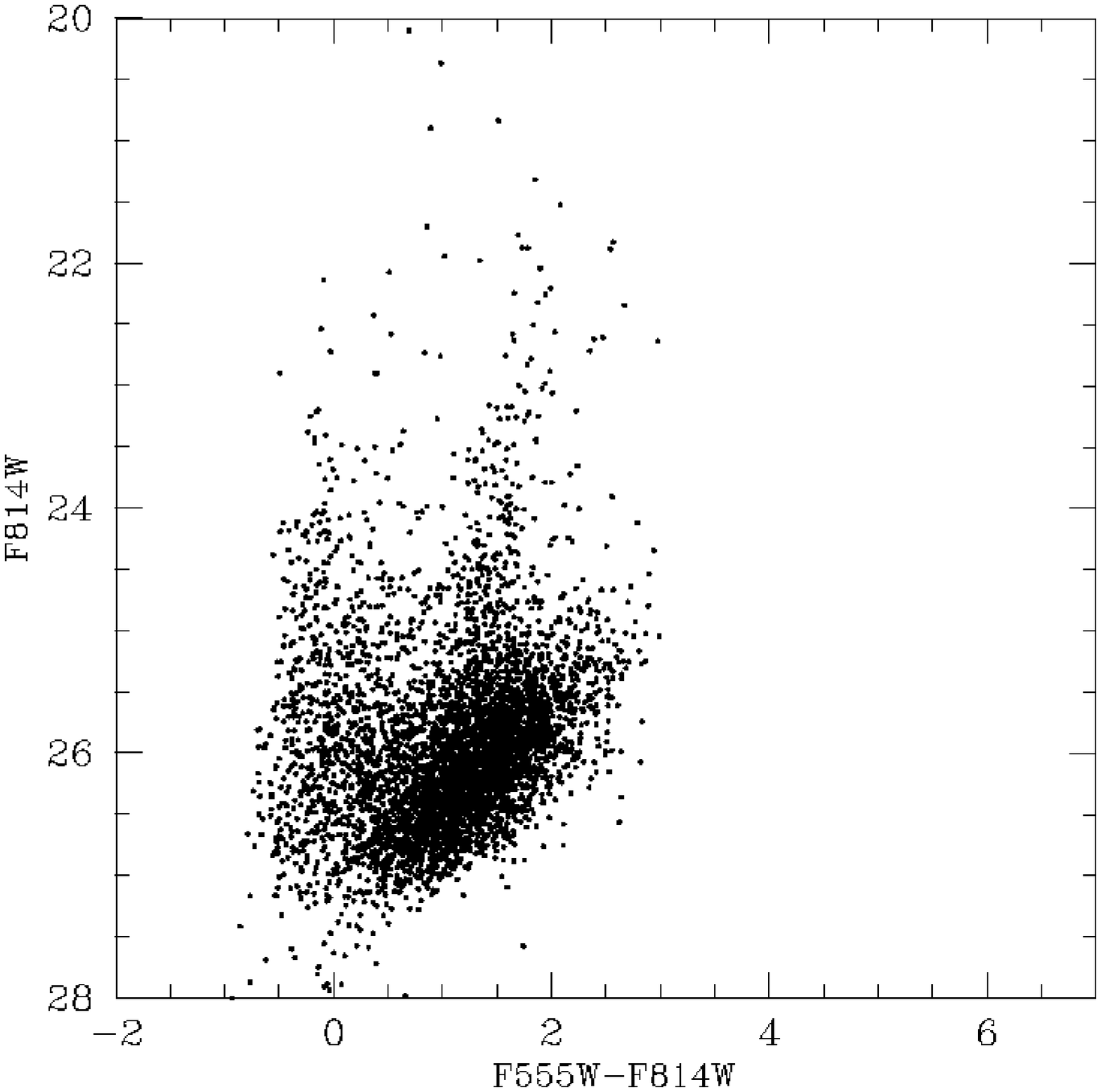}}}
\caption{\label{cmd0} CMD of UGC 5889 (whole field)}
\end{figure}

The CMD of the whole galaxy is
plotted in Fig.\ref{cmd0}.
The blue plume is located at (F555W-F814W)$<$ 0.5 and extends from the
limiting magnitude at F814W $\sim$ 27 up to 22. 
The red evolved stars extend from F814W $\sim 27$ to $\sim 23$ and have
 (F555W-F814W)$>$0.5.

%We used the relations in Holtzman et al. (\cite{holt+96b}) 
%to convert the F555W and 
%F814W magnitudes to standard Johnson/Cousins magnitudes.

%*** check for Cepheids in the galaxy *** 

\section{Distance modulus determination}
\label{distance}
\begin{figure}[bt]
%\vspace{-8mm}
%\special{psfile=figcomp2b.ps
%         hoffset=-30 voffset=00 hscale=37 vscale=37 angle=-90}
%\vspace{53mm}
%{\resizebox{9.0cm}{!}{\includegraphics{z008.ps-f4c.ps}}}
%%{\resizebox{9.0cm}{!}{\includegraphics{figgrad.ps}}}
{\resizebox{9.0cm}{!}{\includegraphics{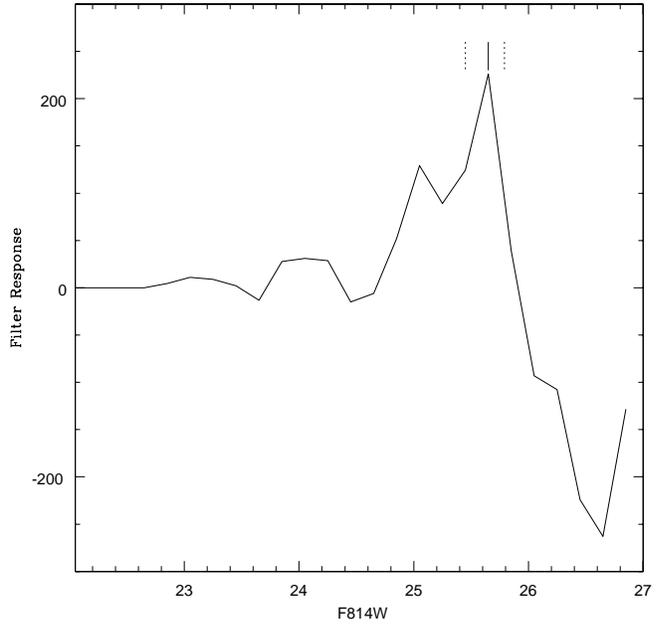}}}
\caption{\label{grad} RGB tip detection: the gradient filter response to the luminosity function is plotted  as a function of the magnitude(see text for more detail). The position of the main peak is taken as indication of the RGB tip (solid line). The dotted lines show the half width at half maximum of the peak taken as estimate of the uncertainties on the tip detection}
\end{figure}

The I magnitude of the
tip of the red giant branch (TRGB) has been shown to
be an excellent distance indicator for nearby resolved galaxies 
(Lee et al. \cite{lee+93}). 
 The position of the TRGB in UGC~5889 
has been obtained dividing the observational
   F814W--(F555W-F814W) CMD into small intervals of width (0.15, 0.15) and then
 convolving the bi--dimensional histogram with an edge detector function,
namely a gradient filter with an angle of 90 degrees, increasing
the contrast at increasing magnitudes. The sharp cut-off at the end of the RGB is clearly detected by the gradient filter (see Fig\ref{grad}).
We take the position of the main peak as our estimate of the tip magnitude and the half width at half maximum of the peak as the uncertainty on the determination. We find F814W=25.65$\pm{0.2}$  as the TRGB magnitude.
Since all the existing calibrations of the TRGB magnitude are
expressed in the Johnson passbands, we first convert the
de-reddened HST F555, F814 magnitudes to the V, I magnitudes. 

%Following the relations of Holtzman et al. (\cite{holt+97}), 
%the F814--(F555-F814) magnitude and colour at the TRGB  de-reddened.

We derive the foreground reddening from the 
maps of Schlegel et al. (\cite{schl+98}), $E_{\rm B-V}=0.034$,
resulting in $E_{\rm F555W-F814W} = 0.06$. This value is slightly lower than
the $E_{\rm B-V}=0.06$ estimated by Knezek et al. (\cite{knez+99}) using
the standard dust-to-gas ratio of Diplas \& Savage (\cite{dipl+94}) and
the H\,I measurements of Hartmann \& Burton (\cite{hart+97}).
The resulting difference in magnitude is within the photometric errors
and hence does not affect  our further conclusions.

We therefore use 
$E_{\rm (B-V)}=0.034$ corresponding
to $E_{\rm (F555-F814)}=0.06$ and $A_{\rm F814}=0.09$.
With these values,
the magnitude and colour at the TRGB transform 
into the Johnson I$_0$=25.47, (V-I)$_0$=1.48.

The distance modulus is derived from
the $I_{TRGB}$ magnitude at the TRGB using the definition:

$$(m-M)_0 = I_{TRGB}-M_{I,TRGB}$$

The dependence of $M_I^{TRGB}$ on the metal content$ [M/H]$ is given by
the recent calibrations of Bellazzini et al.(2001, 2004):

%$$M_I^{TRGB}=0.14[Fe/H]^2+0.48[Fe/H]-3.69$$

 $$M_I^{TRGB}=0.258[M/H]^2+0.676[M/H]-3.626$$

%%and the metal content $[Fe/H]$ is obtained as a function of the 
%%colour of the TRGB from empirical calibrations:

%%%$$(V-I)_0^{TRGB}=0.581[Fe/H]^2+2.472[Fe/H]+4.013$$

This holds true for populations in the age range from 2 to 12 Gyr and 
$ -1.5 <[M/H] <  -0.6$. This method to derive the distance has been
proved to be independent of the details of the star formation history
(Barker et al. 2004). In the above metallicity range, $M_I^{TRGB}=-4.05$
when  $[M/H]=-1.5$ and  $M_I^{TRGB}=-3.91$
when  $[M/H]=-0.6$. 
Since the metal content of UGC~5889 is likely inside the above range (see following Section), we are probably not seriously in error if we
assume  $M_{I,TRGB}=-3.98\pm 0.07$.
%%Using the above relations, we derive the metal content
%%$[Fe/H]=-1.8$ or Z=0.0003, which yields the absolute magnitude $M_{I,TRGB}=-4.05$, 
This results in a  distance modulus of $(m-M)_0=29.45\pm 0.2$, where the quoted errors take into account both the uncertainties on the absolute magnitude  and on the observational magnitude of the tip.
%in agreement with previous determinations.
This value is in reasonable agreement with previous determinations.
For the companion NGC\,3377, Graham et al. (\cite{grah+97}) derive 
a distance modulus  $30.01\pm0.15$. 
A direct determination for UGC\,5889 has been
made by Makarova \& Karachentsev (\cite{maka+98}) using the magnitude of the
brightest blue stars, and they find (m-M)$_0$=29.8.

\section{CMD discussion}
\label{cmd}
\begin{figure}[bt]
%\vspace{-8mm}
%\special{psfile=figcomp2b.ps
%         hoffset=-30 voffset=00 hscale=37 vscale=37 angle=-90}
%\vspace{53mm}
%{\resizebox{9.0cm}{!}{\includegraphics{z008.ps-f4c.ps}}}
%%{\resizebox{9.0cm}{!}{\includegraphics{fig4_i4.ps}}}
{\resizebox{9.0cm}{!}{\includegraphics{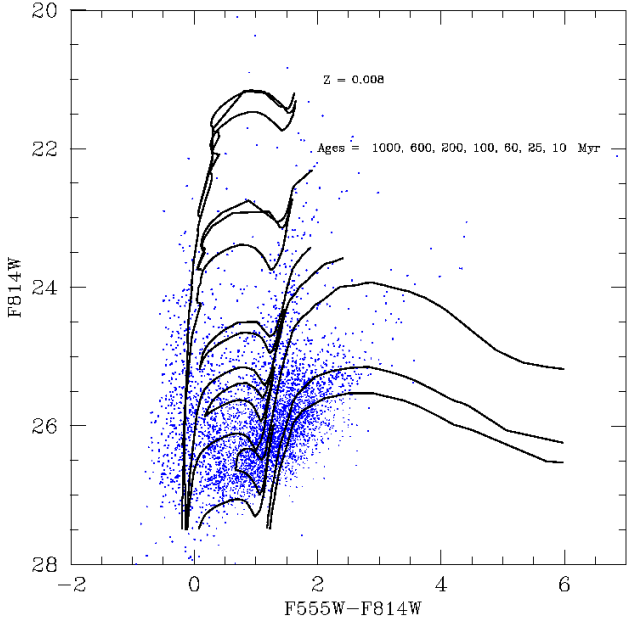}}}
\caption{\label{isoz008} CMD of UGC 5889 with isochrones
of metallicity $Z=0.008$. For  clarity, not all the stars
in the field of UGC 5889 are plotted.}
\end{figure}

To derive  information about the SFR, we compare
the CMD with isochrones of different metallicity.
Fig.\ref{isoz008} shows the CMD  with the isochrones of $Z=0.008$ which 
reproduce the main features of the CMD such as the slope of the giant branch
and the location of those stars brighter than F814W$\sim 26$ and bluer than 
(F555W-F814W)$\sim 0.5$ (blue plume). This blue plume is formed by a
mixture of main sequence and  blue loop stars 
 which crowd the same CMD region.
%Slightly lower values of $Z$ cannot be excluded,
%however 

\begin{figure}[bt]
%\vspace{-8mm}
%\special{psfile=figcomp2b.ps
%         hoffset=-30 voffset=00 hscale=37 vscale=37 angle=-90}
%\vspace{53mm}
%{\resizebox{9.0cm}{!}{\includegraphics{z02.ps-f5.ps}}}
%%%{\resizebox{9.0cm}{!}{\includegraphics{fig5_i4.ps}}}
{\resizebox{9.0cm}{!}{\includegraphics{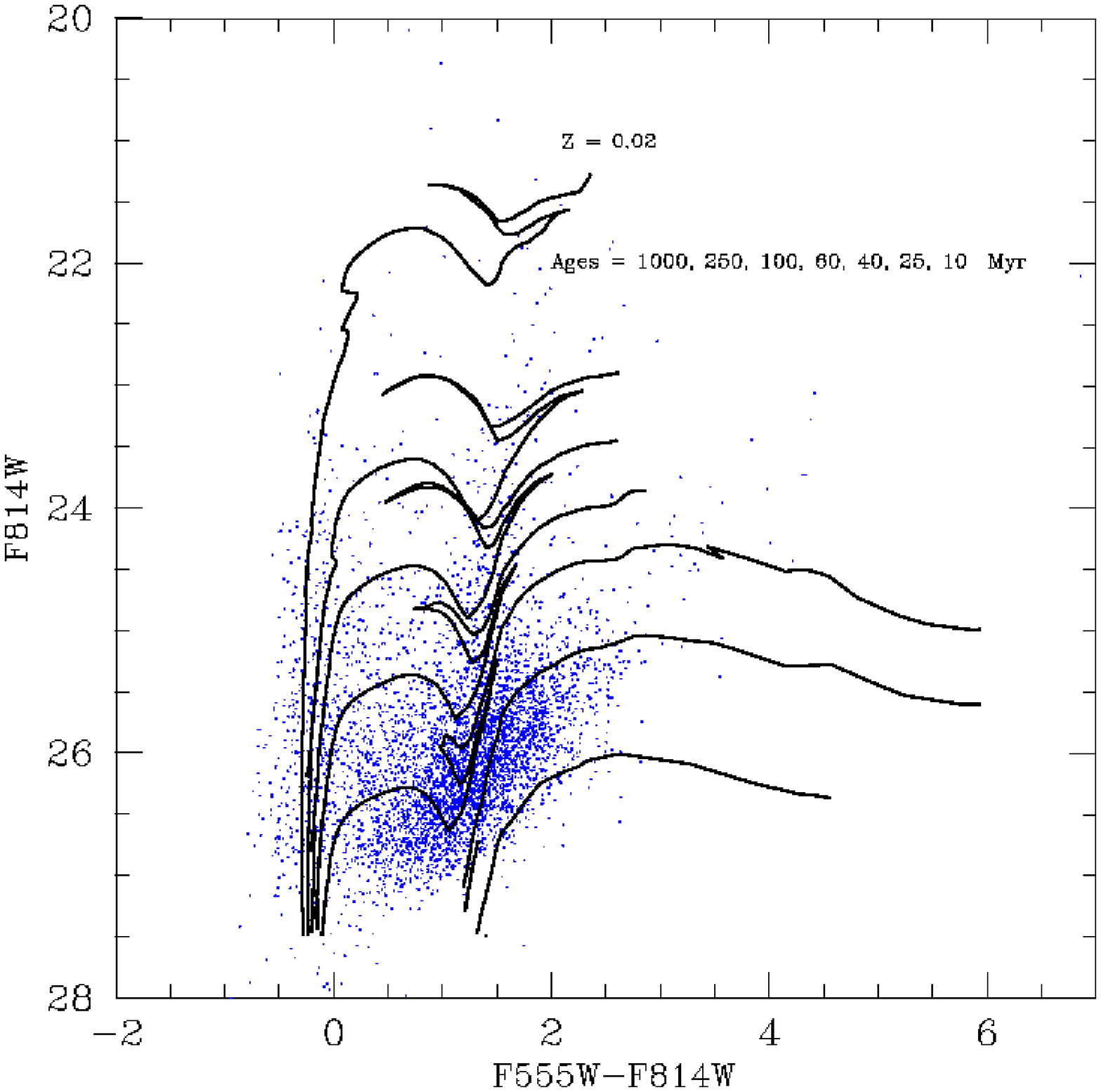}}}
\caption{\label{isoz02}  CMD of UGC 5889 with isochrones
of metallicity $Z=0.02$.}
\end{figure}

A solar metallicity is only marginally consistent, predicting a 
colour of the core He--burning stars redder than observed (see Fig.\ref{isoz02}).
%Assuming a metal content of $Z=0.008$ we find that
%the main sequence
%band bluer than  (F555W-F814W)$ \sim 0.1$  is mainly populated by main sequence stars 
%between 5 Myr and 50 Myr, while a sparse structure at(F555W-F814W)$\sim 0.2-0.5$ and  F814W$ < 25$ is probably due to blue loop stars.
 Core He-burning stars as old as 200 Myr 
are crowding the clump of the red stars brighter than F814W $\sim 27$ 
and having (F555-F814)$ < 1.2$. 
%Finally, AGB stars as old as 600 Myr are expected to be
%brighter than F555W $\sim 27$ and redder than (F555-F814)$\sim 1.2$.
We expect to have a good time resolution for ages younger than about 200 Myr.
At this age, main sequence stars and/or He-burning
loop stars
are brighter than the completeness limit.

\begin{figure}[bt]
%\vspace{-8mm}
%\special{psfile=figcomp2b.ps
%         hoffset=-30 voffset=00 hscale=37 vscale=37 angle=-90}
%\vspace{53mm}
%{\resizebox{9.0cm}{!}{\includegraphics{z0001.ps-f6b.ps}}}
%%%{\resizebox{9.0cm}{!}{\includegraphics{isoz001_i5.ps}}}
{\resizebox{9.0cm}{!}{\includegraphics{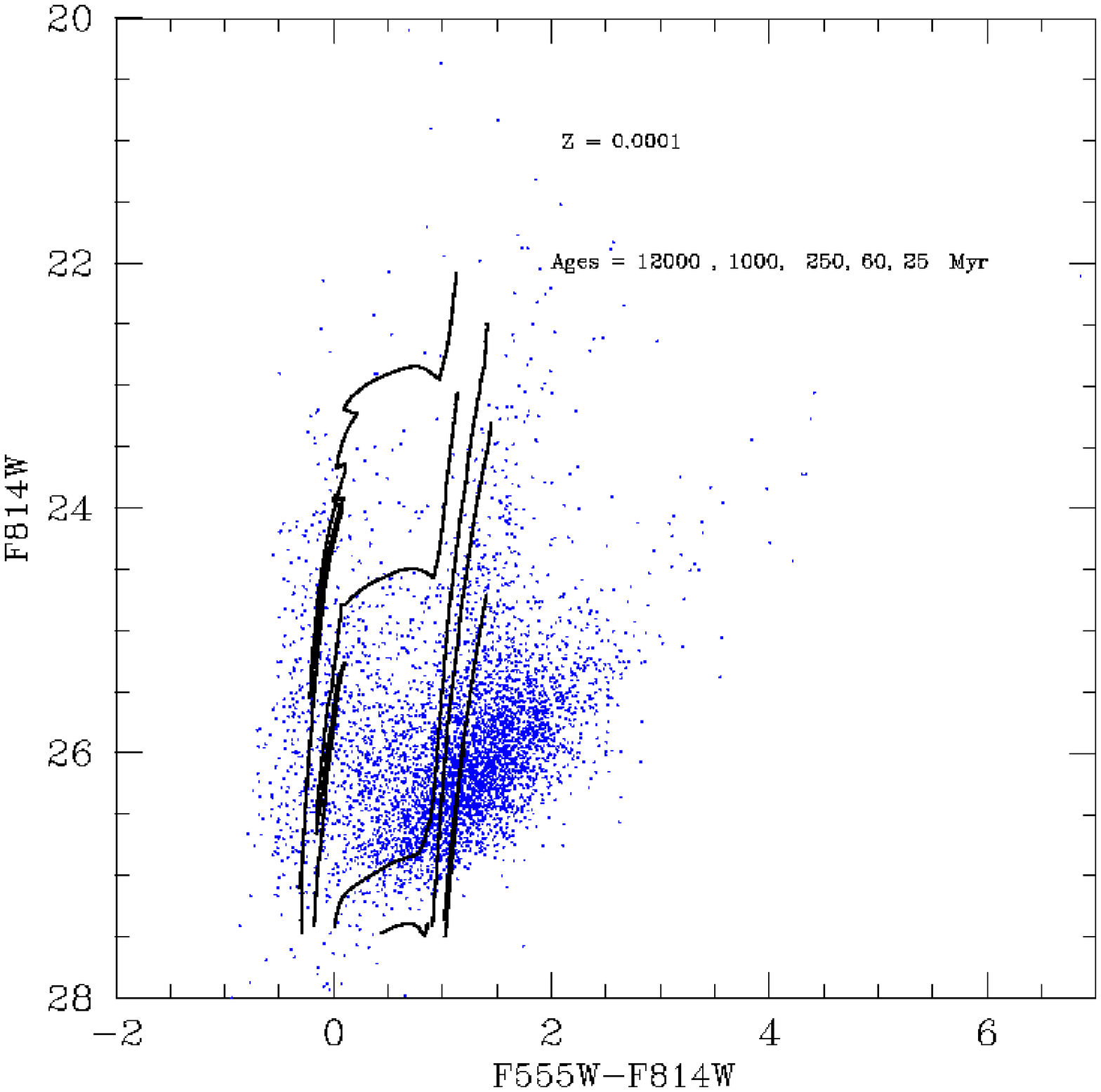}}}
\caption{\label{isoz0001}CMD of UGC 5889 with isochrones
of metallicity $Z=0.0001$.}
\end{figure}

Concerning the presence of an older population, hints can be derived from
the CMD. The large clump redder than (F555-F814)$\sim 1.2$ extending 
down to the photometry limit of F814W $\sim 27$ is consistent with
the presence of AGB and He-burning   stars of $Z=0.008$
in the age range $\sim$ 600-1000 Myr.
Comparing the observational CMD with isochrones as metal-poor as
$Z=0.0001$, it is evident that this metallicity 
cannot reproduce the AGB/RGB colour of stars younger than 1 Gyr
(see Fig.\,\ref{isoz0001}). 
However, the existence of a metal poor population ( $Z \leq 0.0001$) in the age
range 1--12 Gyr cannot be ruled out. Stars of this age in the RGB/AGB phase
are expected to appear at magnitudes fainter than  
F814W $\sim 23.5$ and at a colour (F555-F814)$> 1.2$.

%We point out that  a 
%distance modulus roughly 0.5 nearer would allow to reach the 
%the stars observed at the red tip of the supergiant branch/AGB.
%Comparing the CMD with isochrones, it is quite clear that the
%star formation rate proceeded in this galaxy in a continuous way,
%from about $10^9$ yr till a few $10^6$ yr.

%The distribution of the stars in the red part of the CMD
%allow us to  distinguish a main episode of star formation: using $Z=0.001$
%we date it from
%some Gyr to $10^8$ yr. After that time the SFR run probably at very
%low rate until 2.5 $10^7$ yr ago. A minor episode of star formation
%took place from 2.5 to about 1 $10^7$ yr.
%The decrease of the main sequence at V$\sim -5.7$ is probably due
%to a superposition of  stars of ages from 2.5 - 1 $10^7$ in the phase
%of the blue loop 
%rather than indicating a main sequence turnoff of 6$ 10^6$ yr. (*** to be
%verified with simulations).

\section {Star formation rate}
\label{sfh}
%The study of the star formation rate is done first generating a set of synthetic populations 
%at varying the parameters  age, metallicity range, star formation
%law and initial mass function. Then, a downhill simplex algorithm is
%used to find the SFR that minimize the

\begin{figure}
%{\resizebox{9.0cm}{!}{\includegraphics{cmdbin.ps}}}
%%%{\resizebox{9.0cm}{!}{\includegraphics{fig7_i4.ps}}}
{\resizebox{9.0cm}{!}{\includegraphics{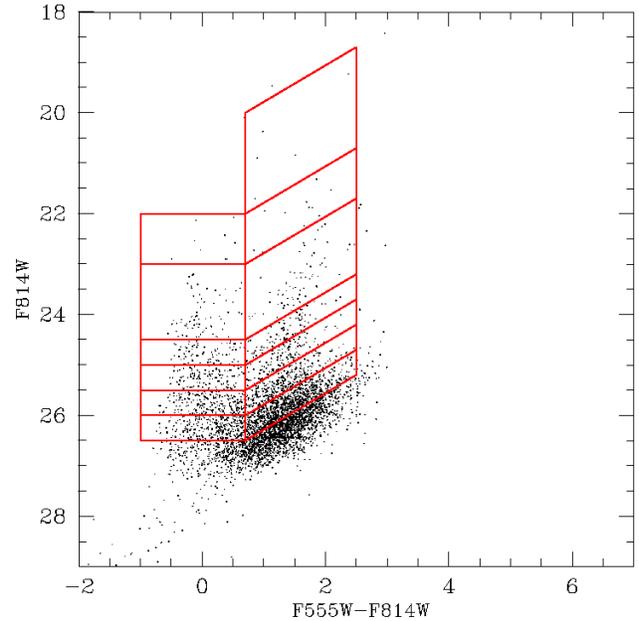}}}
\caption{\label{bins} The subdivision of the CMD into bins.}
\end{figure}

In order to infer the SFR of this galaxy,
theoretical CMDs in different age ranges are simulated. These include 
the spread due to observational photometric errors. 
For each age bin, 1000 stars were generated. 
The generation of the synthetic populations makes use of the
set of stellar tracks by Girardi et al. (\cite{gira+96}; \cite{gira+00}).
%and Bertelli et al. (\cite{bert+94}).
The initial mass function (IMF) of 
Kroupa (2001, 2002) is assumed.
This IMF is a power-law function with a slope $\alpha=2.3$ for stellar
masses m $> 0.5$ M$_{\sun}$, while $\alpha=1.3$ in the mass range 0.08-0.5 M$_{\sun}$ (when the standard Salpeter value is 2.35).
The constancy of the IMF slope in different environments is still a matter of 
discussion, although a number of recent papers  
proposed the idea of a universal IMF 
(see e.g. Kroupa 2002, Wyse et al. 2002, Chabrier 2003, Weidner \& Kroupa 2004).
Recently, it has been suggested that $\alpha$ might depend
on the galaxy type. For the  
IMF in LSB galaxies, it has been proposed that
the effect of the low metal content can result in an inefficient SF 
taking place mainly outside the giant molecular clouds, thus yielding an 
IMF deficient in massive stars.
 As a consequence the IMF slope might be as steep as 
$\alpha \sim 3.8$ (Schombert et al. 1990).  However, in this
paper a more
conservative approach is used, since a variation of the IMF slope in LSB galaxies is not
at present completely justifiable (Lee et al. 2004).
%%%%%biblio da Lee et al
The completeness of the data is taken 
into account by dividing the simulated CMD in magnitude--colour bins
and then subtracting from each bin  having N$_{th}$ stars,
  (1-$\Lambda$)N$_{th}$, where $\Lambda$ is the smallest of the F555W and
F814W completeness
factors as given in Fig.\ref{comple}. 

Finally, the SFR is derived by means of a
downhill simplex method (Harris \& Zaritsky 2001, Rizzi et al. 2002),
minimising the $\chi^2$ function in a 
parameter-space having N dimensions.
At each step the local $\chi^2$ gradient
is derived and a step in the direction of the gradient is taken, until a
minimum is found. 
In the following,
the observational CMD is divided into  bins. Recent work concerning the determination of the
SFR from the CMDs has pointed out the importance of using binning that takes into account the various stellar evolutionary 
phases, as well as the uncertainties on the stellar models (Rizzi et al. 2001).
For this reason, while a coarser magnitude bin distribution is used,
only two bins in colour are considered, namely from -0.5 to 0.5,
corresponding to blue plume stars, and from 0.5 to 3, where the
evolved stars are located. 
This avoids that the uncertainties on both the observational errors
and the theoretical models (i.e. on bolometric corrections, 
RGB and AGB location, extension of the core He-burning loop) 
result in spurious solutions. Additionally the magnitude bins are taken
in the blue part of the CMD at constant F814W magnitude, while in the red part
they are at constant F555W magnitude.
The reason for  this is that the completeness limit in the red part of the 
CMD is mainly dependent on the completeness of the  F555W magnitude.
 This is illustrated in
Fig. \ref{bins}.
To prevent settling on local rather than global minima, the simplex is first
 started
from a random position, then when a possible solution is obtained,
it is re-started from a position very close to it. Finally, when a  minimum
is found,  30000 random directions are searched for a new minimum.

The first guess solution is obtained by comparing the observational CMD with 
isochrones of different ages and metallicities (see previous section).
We use N=14 stellar populations, whose ages
are listed in  Table~\ref{ages}.
The chemical enrichment law is assumed as a pre-defined input parameter,
derived comparing data and isochrones.
Following the discussion of the previous section for ages younger
than 20 Myr, we assume that the stars
are generated with a metal content randomly chosen
in the range Z=0.004-0.01. SFR episodes in the age range 20 Myr-200 Myr 
are assumed to be produced by stars having Z=0.001-0.01.
An older population in the age range  0.2--1 Gyr 
with the same metal content is included.
A population of metal-poor stars with Z=0.0003-0.004 and as old as
 1-12 Gyr is also taken into account, since
due to the low metal content, their magnitudes will be brighter than the
photometry limit of F555W $\sim$ 27.5. Metal-rich stars in the same age range are not considered
since they are expected to be too faint.
As discussed in Section \ref{cmd}, we are aware that we 
have poor age and SF resolution for populations older than 200 Myr.
% Additionally, at magnitudes fainter than F814W $\sim 26.5$
% the completeness correction is higher than
%the  limit of 40\% to be reliable. 
Therefore, the old age bin is included only to
get an indication of the presence of an old population. 

%The minimum $\chi$ solutions 
%are listed in Table \ref{sol} together with
%the  $\chi^2 = 1/\nu \sum _i (a_i-b_i)^2/a_i$ where $a_i$ and $b_i$
%are the observed and the expected number of stars per magnitude bin,
%and $\nu$ is the number of degrees 0.005        0.02          70
%        0.01       0.015           0
%%       0.015        0.02           0
%        0.02        0.03           0
%        0.03        0.04       28.13
%        0.04        0.05           0
%        0.05        0.06           0
%        0.06        0.07       24.37
%        0.07        0.08           0
%        0.08        0.09           0
%        0.09         0.1       8.125
%         0.1         0.2       30.81
%         0.2           1       41.56
%           1          10       1.778
%of freedom (which is equal to the number
%of bins minus 1, since the in the simulations we impose that the
%total number of model stars is equal to the observed total number of stars).

\begin{figure}[bt]
%\vspace{-8mm}
%\special{psfile=figcomp2b.ps
%         hoffset=-30 voffset=00 hscale=37 vscale=37 angle=-90}
%\vspace{53mm}
%{\resizebox{9.0cm}{!}{\includegraphics{f8b.ps}}}
%%%{\resizebox{9.0cm}{!}{\includegraphics{fig8new.ps}}}
{\resizebox{9.0cm}{!}{\includegraphics{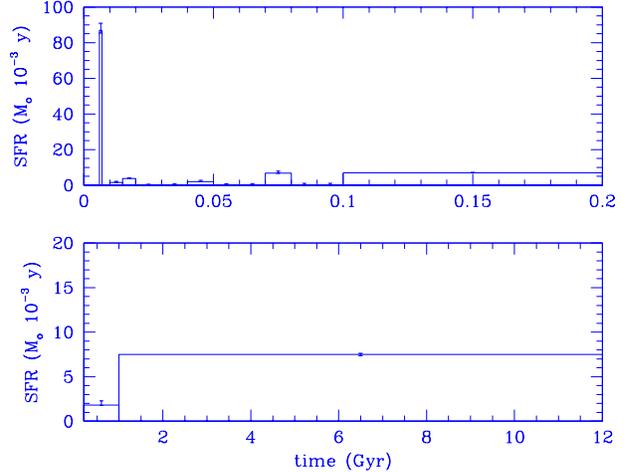}}}
\caption{\label{sfr} SFR of UGC 5889 as a function of look-back time.
Note that due to the low age resolution beyond 1Gyr, we 
only give the average SFR 
in the lower plot (see text and table \ref{ages}).}
\end{figure}
\begin{table}[t!]
\caption{\label{ages} The solution parameter}
\begin{minipage}[t]{8cm}
{\footnotesize 
\setlength{\tabcolsep}{0.15cm}
\begin{tabular}{c|c c c c c}
\hline
\hline
 \noalign{\smallskip}
&$t_{in}$& $t_{fin}$&SFR&$SFR_{low}$&$SFR_{high}$\\
&Gyr&Gyr&10$^{-3}$M$_{\sun}/y$&10$^{-3}$M$_{\sun}/y$&10$^{-3}$M$_{\sun}/y$\\
 \noalign{\smallskip}
\hline
 \noalign{\smallskip}
SFR1&      0.006 &      0.007 &     86.9&       85.3 &       90.9\\
SFR2&        0.010 &      0.015 &      1.6   &    1.6   &    2.1\\
SFR3&       0.015&        0.020 &    3.7   &     3.6   &    4.2 \\
SFR4&        0.020 &       0.030 &      0.0 &      0.0&       0.6\\
SFR5&         0.030 &       0.040&      0.17 &     0.17 &      0.8\\
 SFR6&       0.040 &       0.050 &      2.0 &      1.7&      2.8\\
SFR7&        0.050  &      0.060&     0.0 &   0.0 &      0.9\\
SFR8&        0.060 &       0.070&        0.0 &      0.0 &      1.0\\
SFR9&        0.070 &       0.080 &   6.9 &  6.7 &     7.9\\
SFR10&        0.080 &       0.090 &      0.0 &      0.0  &     1.1\\
SFR11&         0.090 &        0.100 &      0.0&       0.0 &      1.1\\
 SFR12&        0.100  &       0.200 &      7.0 &       6.9  &     7.3\\
 SFR13&        0.200  &         1.000&   1.8 &   1.7  &    2.2\\
 SFR14&          1.000  &        12.000 &     7.5 &      7.3   &    7.6\\
 \noalign{\smallskip}
\hline
\hline
\end{tabular}
}
\end{minipage}
\end{table}

In Fig.\ref{sfr} the best solution is plotted.
The corresponding best fit SFRs
are given in Table~\ref{ages}. 
We derive a $\chi^2$ value of 1.7.
In order to better understand the quality of
the fit, we  compare the values of $\chi^2$
from two simulations with the same observational errors
and input  parameters, but 
different random number generators. The resulting $\chi^2$ for the 
second solution is 1.5 and supports the original finding.
The range of SFR 
corresponding to  68\% of the confidence level  is given
in Table~\ref{ages}. This formal error is obtained 
by exploring the parameter space by varying all the amplitudes
simultaneously in random directions, stepping away from the minimum
until the $\chi^2$ value indicates that we have
reached the 68\% confidence level. The number of random directions
to be checked has been tested on the simulations. About 50000 random directions
are explored and are found to be sufficient to derive the confidence level. 

The recent star formation appears to have proceeded in modest episodes
at a rate of the order of 10$^{-2}$ M$_{\sun}$/yr,  
with periods of  lower star formation or even quiescence.
A very short burst  took place 6 Myr ago and lasted about 1\,Myr.
Any other solution involving different durations, intensities, or age scales
for star formation episodes  younger than 10 Myr
yields a worse fit.

\begin{figure}
%{\resizebox{9.0cm}{!}{\includegraphics{pann1_i4.ps}}}
%%{\resizebox{9.0cm}{!}{\includegraphics{fig9new.ps}}}
{\resizebox{9.0cm}{!}{\includegraphics{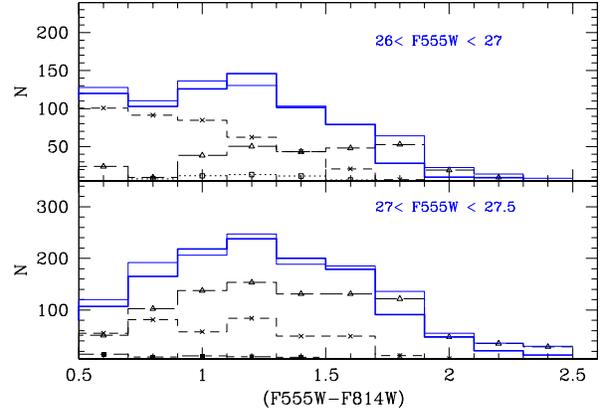}}}
\caption{\label{pann1} The observational colour distribution (heavy solid line)  compared with the simulations (thin solid line) in two magnitude bins  when the population 13 older than 1 Gyr is taken into account. Open squares indicate the simulated colour distribution in the age range 0.09-0.1 Gyr, crosses are the analogous values  in the age range 0.1-0.2 Gyr. Filled squares show the analogous values in the age interval 0.2-1.0 Gyr. Open triangles represent the simulated distribution for ages older than 1 Gyr. }
\end{figure}

\begin{figure}
%{\resizebox{9.0cm}{!}{\includegraphics{pann3_i4.ps}}}
%%{\resizebox{9.0cm}{!}{\includegraphics{fig10_new2.ps}}}
{\resizebox{9.0cm}{!}{\includegraphics{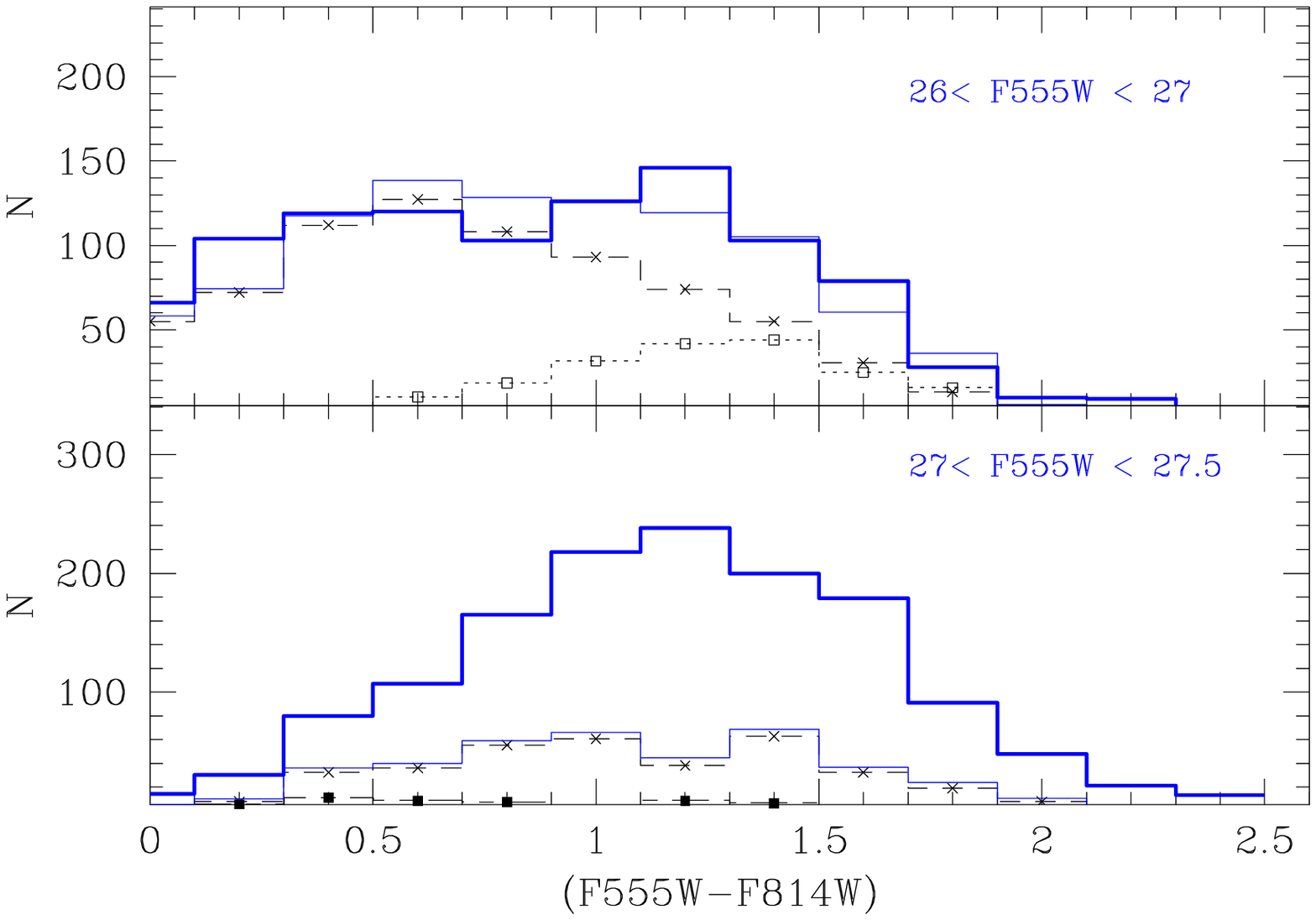}}}
\caption{\label{pann2} The observational colour distribution (heavy solid line)  compared with the simulations (thin solid line) in two magnitude bins  when the population 13 older than 1 Gyr is not included. Open squares indicate the simulated colour distribution in the age range 0.09-0.1 Gyr, crosses are the analogous values  in the age range 0.1-0.2 Gyr. Filled squares show the analogous values in the age interval 0.2-1.0 Gyr. }
\end{figure}

\begin{figure}
%{\resizebox{9.0cm}{!}{\includegraphics{elli_2.ps}}}
%%%{\resizebox{9.0cm}{!}{\includegraphics{ellisse_new.ps}}}
{\resizebox{9.0cm}{!}{\includegraphics{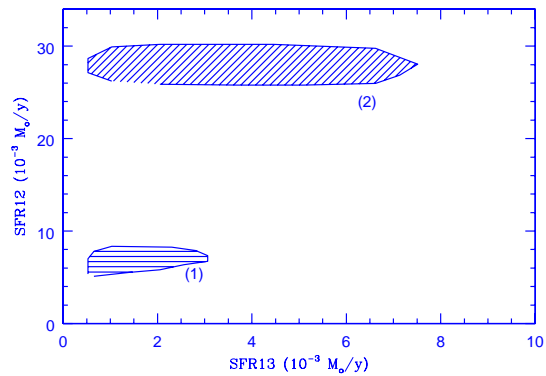}}}
\caption{\label{elli} The confidence interval for the SFR  of population 12 and 13 derived from the red stars  in 
the faintest magnitude bins,
 namely $ 26<$ F555W$< 27$ (1),
$ 27<$ F555W$< 27.5$ (2), when no population older
than 1 Gyr is included in the data (see text for details). }
\end{figure}

To assess the reliability of the solution at old ages, the downhill simplex 
method is applied to a different sets of initial ages, excluding
in turn the oldest populations, in the age range
0.2 - 1 Gyr  and 1-12 Gyr, respectively.  The simplex converges
at high values of $\chi^2$, 4.6 and 2.9 respectively. 
To better verify the presence of an old population, we consider the CMD stars
redder than (F555W-F814W)$\sim 0.5$ and fainter than  F555W$\sim 26$.
This region of the CMD is populated by  stars older than 0.09 Gyr.  
We
analyse the colour distribution in two magnitude bins, namely $ 26<$ F555W$< 27$, $ 27<$ F555W$< 27.5$,corresponding to the two faintest magnitude bins
in the red part of the CMD in Fig.\ref{bins}. We make use of the last 4 populations of Table \ref {ages}.
Figs \ref {pann1} and \ref {pann2} present the best fit solution together
with the data in two cases, a) including the population older than 1 Gyr, and b) without it. In the case b) it is  not possible to fit all the magnitude bins
with the same combination of populations. For instance in Fig. \ref {pann2}
the SFR fitting the bin 27-27.5 cannot reproduce the data in  the brighter
magnitude bins. 
 This is evident as well from Fig.\ref{elli}
where the confidence interval for the SFRs of populations 12 and 13 are plotted  in case b). No common solution is found
 when the population older than 1 Gyr is not included.   
However, taking into account the presence of
observational errors and completeness correction, we prefer to conclude than
the resolution of the data is sufficient
to assess the presence of a consistent population older than 200 Myr, but 
finer age division cannot be reliably derived.
Finally, the stability of the solution against different sub--divisions of
the CMD in cells is tested: slightly larger uncertainties result when a constant bin of 0.5 mag is used.

\begin{figure}

%{\resizebox{9.0cm}{!}{\includegraphics{sim.ps}}}
%%%{\resizebox{9.0cm}{!}{\includegraphics{fig9_i.ps}}}
{\resizebox{9.0cm}{!}{\includegraphics{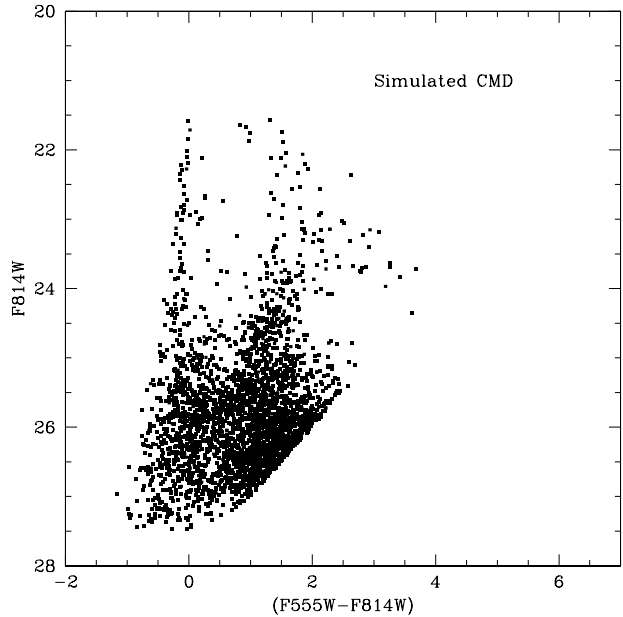}}}
\caption{\label{CMDteo} The simulated CMD using the SFR of Table \ref{ages}
}

\end{figure}

% Fig.\ref{conf}
%presents the confidence interval of 68\%
% for the correlated error of the  oldest populations, in the age range
%0.2 - 1 Gyr and 1-5 Gyr and 5-12 Gyr, respectively.
% The degeneracy of the solutions is evident.
%Unfortunately we cannot answer to the main question concerning
%LSB galaxies, i.e.  whether  they begin to form stars at ages younger
%or older than 5 Gyr.
Fig.\ref{CMDteo} presents the simulated CMD for the best solution.
The simulated CMD is similar to the observational one. 
The total mass going into stars is about 5.5 $\times 10^7$ M$_{\sun}$. 
Since the  total mass of the gas at present is observationally estimated to be
1.6 $\times 10^7$ M$_{\sun}$ 
(Simpson \& Gottesman \cite{simp+00}), the present day 
gas fraction comes out to be 
f$_g$=$M_{gas}/(M_{gas}+M_{stars})=0.2$.  
An independent check of this value can be derived if
f$_g$ is expressed in terms of observable quantities as:

f$_g = (1+ M_{stars}/L_{B}\times L_{B}/ \eta M_{HI})^{-1}$

Bell et al. (2003) derive for a Salpeter IMF: 

$log (M_{stars}/L_{B})= 1.737(B-V)-0.789 $

Using (B-V)=0.64, $M_{HI}/L_{B}=0.3$, and $\eta=1.4$  we find
f$_g=0.17$ in agreement with the determination obtained from the SF.

%\begin{figure}
%{\resizebox{9.0cm}{!}{\includegraphics{conf.ps}}}
%\caption{\label{conf} The confidence interval  of the two oldest populations, in the age range
%0.2 - 1 Gyr (SFR13) and 1-10 Gyr (SFR14)}
%\end{figure}

\subsection {Star formation rate from H$_\alpha$ luminosity}
\label{sfr_alpha}

The total luminosity L$_{H_\alpha}$ emitted in H$_\alpha$ can be used  to estimate the
recent SFR.
Following Kennicutt et al. (\cite{kenn+94}) 
the  $SFR$ can be obtained from 
$$SFR = 1.26 \times 10^{-41} \times L_{H_\alpha}/{10^{38}}~M_\odot yr^{-1}.$$
Adopting L$_{H_\alpha}\sim
2.6 \times 10^{38}$ erg s$^{-1}$  as derived for UGC~5889 by
Knezek et al. \cite{knez+99}, we find
  $SFR \sim 4.0 \times 10^{-3}$~M$_\odot$yr$^{-1}$. 
This SFR should be regarded as the average value over the lifetime
of the stars emitting in ${H_\alpha}$. 
 Assuming that the majority of the
flux in  H$_\alpha$ is due to stars more massive than  10~M$_\odot$,
this gives an estimate of the SFR over the lifetime of a 10~M$_\odot$ star.
This lifetime
can be derived from Girardi et 
al. (\cite{gira+00}) tracks for $Z=0.008$ as 
$20$ Myr.
This SFR can be directly compared with the rate
derived from the star-counts.
Using the coefficients of Table~\ref{ages}, we derive a SFR of $5.6\times 10^{-3}$~M$_\odot$yr$^{-1}$ averaged over the past $20$ Myr of the galaxy
lifetime, in good agreement with the 
L$_{H_\alpha}$ determination.

% The recent star formation is
%much higher than the average over the galaxy lifetime.
%This is not an unusual case among galaxies classified as dwarf Irregulars
%(Gallagher et al. \cite{gall+98}) 
%pointing in the direction of a bursting mode of star formation.

\subsection {The feedback from star formation}  
The HI distribution of UGC~5889 shows a concentration of the gas in a
ring  surrounding
the bulk of the optical body (Simpson \& Gottesman
2000). 
%At the center the gas has  a depression of the global density profile.
% at the center
%UGC~5889 presents a ring shaped distribution of the HI (Simpson \& Gottesman
%2000). This ring appears to be formed by several concentrations, has a depression of the global density profile at the center, and is surrounding
%the bulk of the optical body. 
The velocity field reveals a  rotation of the entire
ring-shaped distribution and no sign of expansion of the ring is found.
This kind of disturbed interstellar medium is not unusual in dwarf
galaxies (Puche \& Westphal 1993, Young \& Lo 1996, Simpson \& Gottesman 2000).
  The hypothesis naturally arises that this dramatic morphology might
be the result of a vigorous star formation episode in a galaxy with a shallow
potential well. 
 In fact the energy injected in the gas by SNs and stellar winds can trigger the
formation of super-bubbles that can easily expand in  a low density medium.
The expansion sweeps up the gas,  forms a cavity at the centre,  
and leaves behind a system where the majority of the gas is located in the
outskirts.
This gas can be lost by the galaxy or can cool and rain back on to
the galaxy after a cooling time (McLow\& Ferrara 1999).
In this section, we discuss whether the energy injected in the gas
by the SN explosions we estimate on the basis of the derived
SFR can be sufficient to explain the morphology of the HI in UGC~5889.
%this morphology can plausibly be driven by
%the observed star formation rate, and
Finally, we  discuss what might be the fate of this gas,
i.e. whether a blow--out or a blow-away can take place.
Following McLow\& Ferrara (1999) and Ferrara \& Tolstoy (2000) we define
a blow--out as a process where the central SN explosions carve a hole in the
gas disk, accelerating a fraction of it, while the remaining gas
stays almost unperturbed. 
In a blow-away nearly all the gas is accelerated
 above the escape velocity and is lost from the galaxy.

To discuss the possibility of blow--out, let us
 assume that, perpendicular to the galactic plane,
the unperturbed gas density distribution in the galaxy is exponential 
$\rho(z)=\rho_0 exp(-z/H)$, where z is the vertical coordinate,
H is the galactic exponential scale height.
The velocity of the shock wave produced by the SN explosions can be expressed as $v \sim (p/\rho)^{(1/2)}$, where p is the pressure.
The velocity $v$ of the shock produced in a stratified medium  decreases
down to a minimum occurring at  z=3H  before being re-accelerated in case of
a blow--out.   
A blow--out takes place when the velocity exceeds the escape velocity of the galaxy at  z=3H, 
where the shock wave is accelerated to infinity.
% the blow--out condition  requires that the blow-out velocity exceeds the escape velocity of the galaxy at a height z=3H, where the curve describing the velocity of the shock
%wave generated by SN explosions has a minimum.
 LSB galaxies are believed to be  dark matter dominated (McGaugh \& de Block 1998).
In order to  calculate the escape velocity, the galaxy is 
 modelled including the presence
of a dark matter halo as in Persic \& Salucci (1996). The dark to visible mass
ratio  $\phi$ can be expressed as a function of the visible galaxy mass 
$M_{g,7}$  in $10^7$ M$_{\sun}$, as

 $\phi \sim 34.7 M_{g,7}^{-0.29}$. 

Assuming as $M_{g,7}=19$ (Simpson \& Gottesman 2000), we find $\phi \sim 14$.
This value is in reasonable agreement with the ratios  in the range 10-30
 derived for LSB galaxies by McGaugh \& Blok (1998).
The blow--out condition can be expressed as (McLow \& Ferrara 1999):

$$ L_{38} > 1.2 \times 10^{-2} M_{g,7} c_{10}^2 h \phi/(\phi+1)$$ 

where $ L_{38}$ is the mechanical luminosity of the SN explosions i.e. the energy produced by SN explosions divided by the average lifetime of the less
massive  star m$_{low}$ exploding as a SN. If we assume
m$_{low}$  $\sim 10 M_{\sun}$ then its lifetime is $20$ Myr.
$ L_{38}$ is expressed in $10^{38}$ erg sec$^{-1}$ when 
$c_{10}=1$ is the sound speed of the interstellar medium in 
10 km sec$^{-1}$ including a turbulence contribution,
and $h$ is the scaled Hubble constant, assumed to be 0.65. 
 Comparing the HI location with the optical image, it is clear that the youngest stars are located in the outskirts of the
gas distribution. Hence, the youngest star formation episode at 5 Myr  
cannot have induced the SN explosions responsible for
 the central gas depletion. 
For this reason, we calculate $ L_{38}$ using 
the mean SFR derived from our model in the last 20 Myr. 
Substituting in the blow-out condition we obtain:

$$ L_{38} \sim 6 >>  0.14 $$.

This means that the blow--out condition holds true inside the gas:
a blow--out is then responsible for the peculiar morphology of the ISM in UGC 5889.
% as it is confirmed by the ISM morphology.
By definition, the blow--out involves only a fraction of the mass of the gas, 
that which is
inside the cavities created by the SN explosions. A more disruptive 
phenomenon is the blow-away in which the gas content of the parent galaxy is
completely lost.
The blow-away can take place when the momentum of the gas shell at the radius
r$_s$ is larger
than the momentum necessary to accelerate the gas outside r$_s$ at a 
velocity larger than the escape velocity. This condition can be expressed following McLow \& Ferrara (1999) as: 

$$ L_{38} > 8 \times 10^{-2} M_{g,7}^{4.9}(\phi/\omega_o)^6  c_{10}^{-10} h$$ 

where $\omega_o=3$ is a scale factor. Under reasonable assumptions concerning
the sound speed $ c_{10}=1.5-0.5$, this condition is not verified even if $\phi$ is assumed to vary
in the range 5-30.
The ultimate fate of the gas seems to be 
that it will not leave its parent galaxy.
% unless
%a new episode of SF is going to inject additional energy in the gas. 
This result is in full agreement with the low gas velocity 
derived observationally (Simpson \& Gottesman 2000).
%On the basis of the above calculations is quite difficult to support the idea that this galaxy is a transition object between a LSB and a dwarf elliptical. 

\section{Comparison with other LSB galaxies}  
\label{comparison}    
Bursty star formation history in LSB galaxies has been suggested in the literature.
We quote, from amongst others, Boissier et al. (\cite{boss+03}) who 
find that it is necessary to include star
formation events such as bursts in the models of LSB galaxy 
evolution to account for the observed scatter in  the Tully--Fisher  and
total mass of the gas--to--luminosity relation.  
A burst scenario is proposed by O'Neil et al. 
(\cite{onei+98}) to explain both the blue colour as well as the disrupted
morphology of most blue LSB galaxies.

The dwarf irregular galaxy IC\,1613 has similar colours (B-V = 0.71)
and its SF has been studied in great detail.
Cole et al. (\cite{cole+99}) find a nearly constant SFR
at a low level of $1.6 \times 10^{-3} \rm M_{\odot} yr^{-1} kpc^{-2}$ over the past 1--300\,Myr. However, due to the large angular size of IC\,1613, they 
only study the very central part of it, which might not be representative
of the SF over the whole galaxy. Note for example that in 
UGC\,5889, the recent SF is concentrated on the outer edge,
while the centre is inhabited by the older population. Skillman et al. 
(\cite{skil+03}) compared it with a field situated at the outer edge of
IC\,1613, but studied the intermediate--age and ancient SFR, starting at
an age of 100\,Myr. A comparison with the CMD of Cole et al. yielded no
significant difference. Although their data are consistent with constant SFR, they  find small bursts in the SFR and a strong enhancement of the SFR 
at intermediate ages. However, no evidence is found for a strong, bursty 
SFR in the recent history. In fact, the CMDs presented by Skillman et al. as
well as the one by Cole et al. lack the bright blue stars that are present
in UGC\,5889 and can only be explained by recent SF. Therefore, although 
both galaxies are classified as blue LSB galaxies they  differ in 
their SFR in the sense that IC\,1613 is dominated by SF at intermediate age,
while UGC\,5889 is dominated by more recent SF. This behaviour is also 
reflected by the total mass--over--luminosity ratio, which is slightly
lower for UGC\,5889.

\section{Conclusions}
\label{conclusions}
 In this paper we have discussed the star formation history of the 
LSB galaxy UGC 5889 on the basis of HST archive
data. 
Our main conclusions can be summarised as the following:

1) the RGB tip is detected at F814W=25.65$\pm 0.2$  implying a distance modulus of $(m-M)_0=29.45\pm 0.2$.
 
2) the recent star formation  is proceeding in modest bursts 
at a rate of the order of 10$^{-2}$ - 10$^{-3}$ M$_{\sun}$/yr. 
This rate is lower than the 5-10 M$_{\sun}$/yr derived for high
surface brightness disk galaxies (Kennicutt 1992), but only slightly larger or 
comparable to the 10$^{-3}$ M$_{\sun}$/yr observed in dwarf irregulars
(Hunter \& Gallagher 1986). The SFR
%in the last 20 Myr 
derived in UGC 5889 from 
%the present study is in agreement with the
H$_\alpha$ emission is in agreement with the results obtained from CMD analysis.

3)
The existence of a consistent population older than 200 Myr and possibly
older than 1 Gyr is suggested by our analysis.  
However, this result might depend on the completeness correction and 
observational errors and needs to be verified with deeper photometry.

4) The present-day total mass in stars is of the order of 8.3 $\times 10^7$ M$_{\sun}$. 
Since the  total mass of the gas has been  estimated as
1.6 $\times 10^7$ M$_{\sun}$, 
the gas fraction is 
f$_g$=0.19.

5) The morphology of the gas inside this galaxy reveals a large cavity,
including the main optical body of the galaxy. 
The gas is mainly located in the external regions.
We have calculated the SFR feed-back on the interstellar gas. We find that 
this impressive morphology is consistent with the hypothesis that
SN explosions  have produced a blow--out of the gas from the internal 
regions  towards the external ones. On the basis
of our calculations, it seems however quite unlikely that the final 
fate of this gas is to leave the galaxy.

%5) UGC 5889 at the mean or even at the present rate of star formation
%will not get rid of its gas, and thus it is unlikely to evolve 
%into a dwarf elliptical. We therefore confirm the reflection of Knezek et al. 
%(\cite{knez+99}) that it does not represent a missing link between the two
%galaxy populations.

\acknowledgement{This research has 
made thorough use of the Simbad database operated at CDS, Strasbourg, France. 
This work
was done while AV was visiting scientist at ESO, Santiago. 
We thank the referee, P. Knezek, whose comments greatly improved this paper.
 We thank K. O'Brien for a careful reading of the manuscript. We are indebt to G. Bertelli for many fruitful discussions.}

\end{document}